# Generating Domain-Specific Transformation Languages for Component & Connector Architecture Descriptions

Lars Hermerschmidt, Katrin Hölldobler, Bernhard Rumpe, and Andreas Wortmann
Software Engineering, RWTH Aachen University, http://www.se-rwth.de/

*Abstract*—Component-based software engineering (CBSE) decomposes complex systems into reusable components. Model-driven engineering (MDE) aims to abstract from complexities by lifting abstract models to primary development artifacts. Component and connector architecture description languages (ADLs) combine CBSE and MDE to describe software systems as hierarchies of component models. Using models as development artifacts is accompanied with the need to evolve, maintain and refactor those models, which can be achieved by model transformations. Domain-specific transformation languages (DSTLs) are tailored to a specific modeling language as the modeling language's concrete syntax is used to describe transformations. To automate the development of DSTLs for ADLs, we present a framework to systematically derive such languages from domain-specific C&C language grammars. These DSTLs enable to describe such model transformations concisely in vocabulary of the underlying ADL. These domain-specific transformations are better comprehensible to ADL experts than generic transformations.

## I. MOTIVATION AND PROBLEM STATEMENT

Engineering non-trivial software systems demands techniques to reduce development effort. Component-based software engineering (CBSE) aims to reduce complexity by composing systems from reusable components. Ideally, these components can be developed independently by domain experts and reused off-the-shelf - increasing component maturity along the way. Components of CBSE usually are source code artifacts, which gives rise to "accidental complexities" [1] (dealing with programming instead of domain issues). Model-driven engineering (MDE) aims to abstract from these by lifting abstract models to primary development artifacts. Such models are typically formulated in terms of a domain-specific language (DSL) that reduces noise and trades expressiveness for comprehensibility. In addition, such models can be better reusable, analyzable, and automatically transformable into executable systems. Component and connector (C&C) architecture description languages (ADLs) [2] combine CBSE and MDE to model systems as hierarchies of components.

Using ADLs in MDE gives rise to needs for multiple types of model transformations, such as: i) preprocessing: translate ADL keywords into equivalent component structures or flatten the component hierarchy prior to code generation, rearrange the subcomponent hierarchy for deployment. ii) refactoring: find architectural anti patterns and replace these with established solutions. iii) refinement: replace platform-independent with platform-specific components.

Describing transformations either requires handcrafting code to transform a model based on its representation, such as an abstract syntax tree (AST), in a general purpose programming language or modeling with a generic transformation language such as ATL [3]. The former is tedious and error prone. The latter requires learning a new language, which might provide adequate transformation descriptions, but cannot rely on the original DSL's notations.

Domain-specific transformation languages (DSTLs) also called "transformations in concrete syntax" [4]–[7] reduce the effort of learning a transformation language as they employ the familiar DSL's syntax. In addition they allow a more concise definition of transformations as the AST is not involved. Producing such DSTLs however requires the same effort as developing a DSL. To approach this, we have developed a framework to generate DSTLs from DSLs while retaining their vocabulary. With this framework, developers can efficiently describe model transformations in well-known form and the overhead of learning additional modeling elements is minimized.

In the following, Sect. II presents the language workbench MontiCore on which our framework, and the ADLs we generate DSTLs for, build. Afterwards, Sect. III describes the framework before Sect. IV illustrates the resulting DSTLs and their application. Sect. V presents related work. Finally, Sect. VI discusses the approach and Sect. VII concludes.

## II. PRELIMINARIES

The DSTL generation framework relies on the language development and integration mechanisms of the language workbench MontiCore [8]. With this, it parses the grammars of MontiCore DSLs and generates domain-specific transformation languages. MontiCore provides a language to describe the integrated concrete and abstract syntax of DSLs in terms of context-free grammars and means to generate model processing infrastructure, such as tools to parse textual models into an abstract syntax tree (AST), frameworks for language integration and well-formedness checking [9], as well as code generation [10]. Language integration enables aggregation, inheritance, and embedding between DSLs. For

K. Hölldobler is supported by the DFG GK/1298 AlgoSyn.

the latter, the host DSL provides extension points filled by modeling elements of the embedded DSL.

We apply our approach to MontiArc [11], a C&C ADL build with MonitCore and its extension MontiSecArc. Both describe logically distributed software architectures as hierarchies of connected components. Components are black-boxes with interfaces of typed, directed ports. The behavior of atomic components is defined by source code artifacts and the behavior of composed components emerges from their subcomponents. MontiSecArc introduces the *trust level* to distinguish components that might be influenced by an adversary from those which are not that easy to reach. qAs modeling something unknown like an adversary is hard, the trust level describes (physical) protection measures which hinder an adversary to compromise a component. The trust level abstracts from individual measures like locked doors, fences, and video surveillance to focus the model on IT security. A subcomponent's trust level is denoted relative to its containing component and the surrounding of a system is assumed as insecure and hence has the trust level $-1$.

A classical measure to hinder adversaries to access a resource is access control, such as role based access control, or access control lists (ACLs) [12], which is noted by the keyword *access* in MontiSecArc. Access is limited to certain policies, such as roles or ACLs, for specific incoming ports or complete components, where the later is equivalent to access control for all incoming ports of the component. Assigning users to roles or ACLs is left to run-time, such that new users' access rights are defined by the access policy.

To avoid naming problems with policies when composing components, policies of different components are independent, even if they have the same name. When interconnecting components, e.g., client and server as depicted in Fig. 4, where one role has access to different components an *identity* link connects these components. When interconnecting identities, the process of authentication, where a user from a proving component claims to have a role at the verifying component, ensures that only users, which possess this role are able to claim it. In Fig. 4 `Client'` is the proving component and `Server'` the verifying one. To specify proving and verifying component, the `identity` link is directed from the former to the latter.

## III. DSTL Generation Framework

The DSTL generation framework is able to create DSTLs that, in conjunction with additional generated and provided parts of the framework, realizes a graph transformation approach. In such approaches, complex transformations are composed of small transformation rules where transformation rules usually are described by a left-hand side (LHS) - the model part before applying the rule - and a right-hand side (RHS) - the same model part after being transformed [13]. The following sections explain the framework to automatically derive DSTLs from the grammar of a modeling language as well as the resulting DSTL and the application of domain-specific transformations.

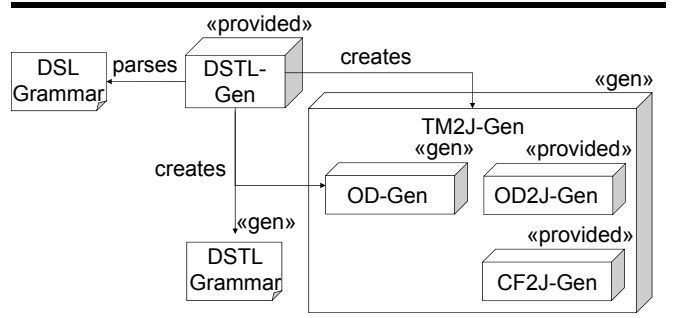

Fig. 1. Overview of the generation of a DSTL including provided and generated generators.

### A. From DSL to DSTL

The automatic derivation of DSTL is solely based on the grammar of the DSL. Thus, by taking the modeling language's grammar as input the DSTL generator produces the grammar for describing transformation rules following the derivation rules described in [14] and a generator (`OD-Gen`, Fig. 1) to translate those transformation rules to a LHS and a RHS of a transformation in form of object diagrams (OD notation). Furthermore, the framework provides a generator to translate this OD notation to Java (`OD2J-Gen`), a control flow language to control the application of transformation rules and a generator (`CF2J-Gen`) to translate the control flow to Java.

As complex transformations usually are decomposed to transformation rules combined by some kind of application strategy ([15]), the control flow language uses the transformation rule language via language embedding to allow the description of complex transformations in form of so called transformation modules (Sect. III-B4). Finally, to ease the use of the generators (`OD-`, `CF2J-`, and `OD2J-Gen`) the DSTL generator creates glue code that combines those generators to a single generator (`TM2J-Gen`) able to translate transformation modules to executable Java transformations.

### B. Generated DSTLs

A DSTL created by the generator described above reuses the concrete syntax of the DSL to describe patterns. In addition, the DSTL provides a replacement operator for modifications, allows to bind elements to variables, and to specify negative elements and application constraints. With this, the DSTL is able to describe endogenous in-place transformations [15], [16]. In contrast to the typical transformation form consisting of LHS and RHS, we use an integrated notation of LHS and RHS. Combining these in a single model avoids repeating unchanged model parts on the RHS. The transformation operators, such as the replacement operator or negative elements, are provided for every model element defined by a nonterminal such as components and ports. The following explains those operators.

*1) Pattern and Schema Variables:* The DSTL uses concrete DSL syntax to describe patterns, thus, a pattern resembles the model part it describes and omits parts that do not constrain the pattern. For example, the model in Lst. 1 could also serve as a pattern. However, every component that has the depicted structure and arbitrary additional structures, such as additional ports or subcomponents, would be a suitable match for this pattern. There also is no need to start a pattern at the top-level element of a model. Instead, all elements can be top-level elements in a pattern. For instance, if a transformation is defined for a port and the containing component is irrelevant, the pattern may only define the port and its modification.

In many cases transformations need to be more general, thus, for abstraction purposes as well as binding model elements to variables (for instance to move them), the generated DSTL provides a concept called *schema variables*. Those variables consist of a type, i.e., the name of the nonterminal that defines the model element and a name starting with a $-sign. There are black box and a white box schema variables: Black box variables end with a semicolon ("*ElementType SchemaVar* ;"), while white box variables allow to define the element's structure within double square brackets ("*ElementType SchemaVar* [[ *Element* ]]"). An example black box variable is depicted in line 6 of Lst. 3 for an `access` definition. Line 9-12 of Lst. 3 show a white box variable for a component.

To ease the use of variables for names the type `Name` can be omitted. A schema variable for a name is displayed in Lst. 2 ($name in l. 7, $sp in l. 8). If a schema variable is used for a model element the corresponding element is bound to this variable during pattern matching. Thus, using the same variable twice refers to the same model element in both cases. However, for names we relaxed this such that two occurrences of a schema variable for a name require equality instead of identity. When using variables for abstraction, without the need for referencing them later, the anonymous $_ variable may be employed. It does not bind the model element and, hence, two occurrences neither require identity nor equality (Lst. 2, l. 9).

*2) Modifications:* The generated DSTL uses an integrated notation of the LHS and RHS of a transformation rule. To achieve this the DSTL provides the replacement operator :- that acts on element level ("[[ *Element*? :- *Element*? ]]"). The element left of :- is replaced by the one right of it. If the LHS is left blank an element is created and added. Leaving the RHS blank deletes an element. A modification is illustrated in line 6 of Lst. 2.

*3) Negative Elements, Application Constraints and Assignments:* Negative application conditions [17] are provided in form of negative elements with the following syntax: `not` [[ *Element* ]]. A negative element is an element that must not occur in the model. Furthermore, a *where*-block is provided that allows formulating application constraints and assignments of schema variables. The *where*-block is structured as follows:
   where { *Assignment*∗ *BooleanExpression*? }
It starts with the assignment of schema variables that are not assigned during pattern matching (i.e., parts of the RHS of a transformation). Within the *BooleanExpression* the elements of the transformation bound to schema variables can be used to formulate the constraint. Thereby, the signature of the abstract syntax of the model elements can be used as well as any static Java method. Listing 2 shows a negative element (l. 9) and a *where*-block (l. 11). An example of an application constraint is shown in line 14 of Lst. 3. A transformation will only be applied if all positive elements are found, no match for the negative elements is possible and the application constraint holds.

*4) Transformation Modules:* To control and combine the transformation rules to transformation modules, the generated DSTL is combined with a generic control flow language via language embedding.

A transformation module, as shown in Lst. 2, consists of instructions and transformation methods (introduced by the keyword `transformation`) where the body of a transformation method is a transformation rule. The instruction methods define the application order of transformation methods. The instruction method `main()` is the starting point of a transformation module. Within instructions, Java syntax extended by a special `loop` statement can be used to specify the control flow in an imperative manner. The loop statement applies the following transformation rule until no further match for the pattern can be found.

## C. Translation and Application of a Transformation

A transformation module is defined using the control flow language and its embedded transformation rule language (Fig. 2), which have to be translated to Java code for execution. This translation is performed by the composed and generated generator `TM2J-Gen` (Fig. 1 and Fig. 2). `TM2J-Gen` takes a transformation module as input and internally uses its three subgenerators to translate it to an executable Java transformation. The latter reads a model and applies the transformation described by the transformation module.

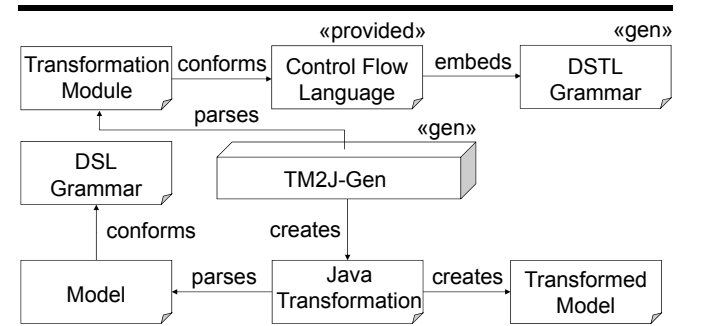

Fig. 2. Overview of the translation and application of a transformation.

## IV. APPLYING THE TRANSFORMATION LANGUAGE

With the DSTL derived from the DSL's grammar, the description of model transformations is greatly facilitated as the transformation developers are familiar with the DSLTs vocabulary. The following sections illustrate application of model transformations to MontiArc and MontiSecArc with the DSTLs generated for each.

### A. Preprocessing: Adding Structural C&C Elements

A common challenge for the development of distributed systems is dealing with the unforeseeable run-time issues.

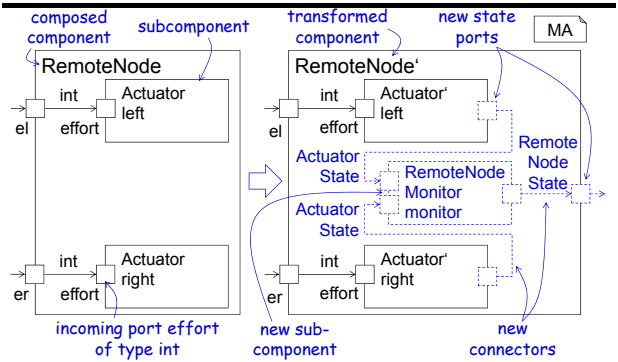

Fig. 3. Applying the monitoring transformation to a composed component RemoteNode with a single subcomponent.

```
component RemoteNode {                                    MA
  port in int el, in int er;
  component Actuator left, right;
  connect el -> left.effort;
  connect er -> right.effort;
}
```

Listing 1. Textual syntax of composed component RemoteNode with subcomponent Actuator.

To this effect, MontiArc introduces component monitoring. Every component of the architecture is monitored by a specific monitor per composed component. Instead of handcrafting the monitoring infrastructure for each component, it is conveniently integrated via model transformations. These transformations introduce new subcomponents, ports, and connectors, such that all composed components and their direct subcomponents are observed by a new subcomponent. That subcomponent receives status messages from its neighboring subcomponents, calculates an overall component status, and emits this via a new outgoing port. Applying this transformation to a composed component RemoteNode (Fig. 3) requires that (a) RemoteNode and all its subcomponents receive a new port to emit status messages, (b) RemoteNode receives a new subcomponent of type RemoteNodeMonitor that provides appropriate input ports for all new state ports and emits messages on the overall state of RemoteNode, and (c) the state ports of the subcomponents of RemoteNode are connected to RemoteNodeMonitor, which itself is connected to the new state port of RemoteNode.

As the DSTL's syntax is derived from the DSL, Lst. 1 describes the textual syntax of the untransformed MontiArc component RemoteNode for comprehension. The keyword component (l. 1), followed by a name and curly brackets declares a component definition (ll. 1-6). The components interface is defined by the keyword port and a list of directed, typed ports (l. 2). Furthermore, a composed component contains a set of subcomponents (l. 3), each starting with the keyword component, followed by its type and name. The ports of subcomponents are connected via unidirectional connectors (ll. 4-5).

Handcrafting these transformations in terms of AST API calls requires considerable effort. Instead, the three transformation rules given in Lst. 2 describe this

```
module AddMonitoring {                                    MTF
  main() { loop addPorts();
           loop addMonitor();
           loop connect(); }

  transformation addPorts() {
    component $name {
      port [[ :- out $sp state ]] ;
      not [[ out $_ state ]]
    }
    where { $sp = $name.concat("State"); }
  }

  transformation addMonitor(){
    component $name {
      [[ :- component $type monitor;]]
      not [[ component $_ monitor; ]]
      [[ :- connect monitor.state -> state; ]];
      component $_ {}
    }
    where { $type = $name.concat("Monitor") }
  }

  transformation connect(){
    component $_ {
      component $type $name;
      [[ :- connect $name.state -> monitor.$sp; ]];
      not [[connect $name.state -> monitor.$_;]]
    }
    where {$sp = $name.concat("State");}
  }
}
```

Listing 2. The transformations required to add a monitor, related ports, and connectors to a software architecture.

transformation. The main block (ll. 2-4) invokes the three transformations addPorts(), addMonitor(), and connect(), where addPorts() (ll. 6-12) adds state ports to all components of the software architecture. To this effect, it iterates over all components (denoted by concrete MontiArc syntax component followed by a name $name) and adds a new outgoing port state to each of the component's ports rule (l. 8), where no such port already exists (l. 9). The port's type is defined by $sp as calculated by the where-block (l. 11). The transformation addMonitor() (ll. 14-22) adds a new subcomponent monitor (l. 16) to each composed component - enforced by requiring that the component contains a subcomponent (l. 19) - that does not already contain a monitor (l. 17). The type of monitor is calculated via $type (l. 21). Finally, the transformation connect() (ll. 24-31) adds new connectors to each composed component to connect its subcomponents to its new monitor. This is better comprehensible than a lengthy program exploiting the AST API and less susceptible to errors arising from accidental complexities of AST programming.

### B. Refactoring: Resolving Anti-Patterns

Architects need to consider security as one out of many nonfunctional requirements. There are numerous commonly known anti-patterns and design flaws [18]. We consider the anti-pattern of client-side authentication [19, p. 687], which is depicted in Fig. 4 using the MontiSecArc language. In this case a client with low trust level enforces access control and a server which has a higher trust level relies on that client. Hence, an attacker able to impersonate the client can bypass access control and compromise the server, as it relies on the client.

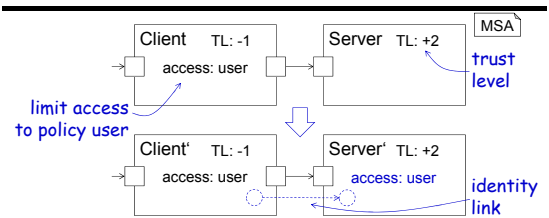

Fig. 4. Applying the transformation to an insecure client server setup introduces access control to the server to make it secure.

```
1  module ClientAuth {
2    main(){ loop accessPort(); }
3
4    transformation accessPort() {
5      SecArcComponent $C [[ component $client {
6        Access $A;
7      } ]]
8      connect $client.$_ -> $server.$someInPort;
9      SecArcComponent $S [[component $server {
10       port in $someInPort;
11       [[ :- access $someInPort ($policy) ]];
12     } ]]
13     where {$policy = $A.getPolicy();
14     $C.getTrustlevel() < $S.getTrustlevel()}
15   }
16 }
```

Listing 3. The transformation moves access control enforcement from client to server components.

We use the transformation depicted in Lst. 3 to identify client components which have this anti-pattern (ll. 5-7) and add access control to the server (ll. 9-12). Client components within this anti-pattern have one of the statement starting with `access`, so we use a black box schema variable `$A` for the common super type `Access` to match both (l. 6). Furthermore, there is a connection (l. 8) to a more trustworthy server (l. 9-12). We use the white box variant of schema variables for the client (`$C` in l. 5) and the server (`$S` in l. 9). In the `where`-block we first retrieve the access policy from the client by utilizing the method `getPolicy()` from `$A` and assign it to the `$policy` variable (l. 13). Finally, to ensure that the client has a lower trust level then the server, we use another method `getTrustlevel()` accessible via the variables `$C` and `$S` (l. 14). Using a combination of keywords and abstract syntax of MontiSecArc in the DSTL makes the patterns precise and comprehensible to domain experts.

## V. RELATED WORK

Similar to PROGRES [20], Fujaba [21], eMoflon [22], and Henshin [23], the transformations of our approach are endogenous, and in-place [16]. However, these approaches do not employ the concrete syntax of the underlying DSL. There are approaches for transforming software architectures [24]–[26], however, they either introduce their own notation, operate on the abstract syntax or provides less functionality e.g. do not allows to remove elements [25]. Existing approaches to derive DSTLs from DSLs focus graphical languages [6], [27] and do not provide the concrete syntax of the transformation language. Another approach to circumvent generic transformation languages is to infer LHS and RHS of model transformations from examples [28], [29]. To generalize these examples, developers have to use abstract syntax. Term rewriting [4] works on concrete syntax as well by applying rewriting rules to manipulate rather small connected model parts as compared to graph transformations. T-Core [30] and others [31] introduce transformation primitives which, similar to term rewriting, do not automate the process of deriving an DSTL but are combined and configured to create it. Thus, they do not propose a systematic and automated way of deriving a DSTL, but provide building blocks to create them. Our previous work on delta languages, which describes small changes for models in concrete syntax of the modeling language, shares the underlying generative approach of deriving those languages we use here and we first applied those deltas to architectural models [32].

## VI. DISCUSSION

A generated DSTL relies on the concrete syntax of its base DSL. However, for typing schema variables the nonterminal names of the modeling language are used and, thus, this abstract syntax information become part of the concrete syntax of the DSTL. This cannot be avoided completely as for the black box variant of schema variables the type cannot be inferred whenever there is an alternative of nonterminals in the base DSL. Furthermore, keywords such as `not` or `where` and delimiters might conflict with the DSL's concrete syntax. However, these problems can be solved by using MontiCore's language inheritance to redefine the concrete syntax of the DSTL. Allowing every model element of the base DSL as a top level element in transformation rules leads to problems if the model element does not have any mandatory concrete syntax. Restructuring the DSL will solve this issue.

## VII. SUMMARY

We presented a framework to generate DSTLs from the grammars of DSLs. The resulting DSTLs consist of declarative transformation rules that employ patterns based on the DSL's concrete syntax to describe both what is to be replaced and how it is to be replaced. These transformation rules are embedded into a control flow language to describe complex, imperative transformation modules. As such DSTLs reuse the well-know vocabulary of the underlying DSL, modeling individual transformations require less effort from domain experts. The control flow language is very compact and learning their combination is less complex than learning a general transformation language. The framework has been applied to the C&C ADLs MontiArc and its descendant MontiSecArc. We currently examine the application of the DSTL generation framework to other ADLs in ongoing case studies.